\documentclass[prl,showpacs,twocolumn]{revtex4}
\usepackage{graphicx}
\begin{document}
\title
{Soliton-dynamical approach to a noisy Ginzburg-Landau model}
\author{Hans C. Fogedby}
\email{fogedby@phys.au.dk}
\affiliation
{Institute of Physics and
Astronomy,
University of Aarhus, DK-8000, Aarhus C, Denmark\\
and
\\
NORDITA, Blegdamsvej 17, DK-2100, Copenhagen {\O}, Denmark
}
\author{John Hertz}
\email{hertz@nordita.dk}
\affiliation
{
NORDITA, Blegdamsvej 17,\\ DK-2100, Copenhagen {\O}, Denmark
}
\author{Axel Svane}
\email{svane@phys.au.dk}
\affiliation
{
Institute of Physics and Astronomy,
University of Aarhus,\\
 DK-8000, Aarhus C, Denmark
}
\date{\today}
\begin{abstract}
We present a dynamical description and analysis of non-equilibrium
transitions in the noisy Ginzburg-Landau equation based on a
canonical phase space formulation. The transition pathways are
characterized by nucleation and subsequent propagation of domain
walls or solitons. We also evaluate the Arrhenius factor in terms
of an associated action and find good agreement with recent
numerical optimization studies.
\end{abstract}
\maketitle
Phenomena far from equilibrium are widespread including turbulence
in fluids, interface and growth problems, chemical and biological
systems, and problems in material science and nanophysics. Here
the dynamics of complex systems driven by weak noise,
corresponding to rare events, is of particular interest in the
context of e.g., nucleation during phase transitions, chemical
reactions, and conformational changes in macromolecules. The weak
noise limit is associated with a long time scale corresponding to
the separation in energy scales of the thermal energy and the
energy barriers between metastable states; the transition takes
place by sudden jumps between metastable states followed by long
waiting times in the vicinity of the states. The fundamental issue
is thus the determination of the transition pathways and the
associated transition rates.

A particularly interesting non equilibrium problem of relevance in
the nanophysics of switches is the influence of thermal noise on
two-level systems with spatial degrees of freedom, see
\cite{Berkov98,Koch00,Garcia99}. In a recent paper by E, Ren, and
Vanden-Eijden \cite{E02} this problem has been addressed using the
Ginzburg-Landau equation driven by thermal noise. These authors
implement a powerful numerical optimization technique for the
determination of the space time configuration minimizing the
Freidlin-Wentzell \cite{Freidlin98} action and in this way
determine the orbits and their associated action yielding the
switching probabilities in the long time-low temperature limit.
The picture that emerges from this numerical study is that of
noise-induced nucleation and subsequent propagation of domain
walls across the sample yielding the switch between the two
metastable states.

In recent work we have addressed a related problem in
nonequilibrium physics, namely the Kardar-Parisi-Zhang equation or
the equivalent noisy Burgers equation describing for example a
growing interface in a random environment. Using a canonical phase
space method derived from the weak noise limit of the
Martin-Siggia-Rose functional \cite{Martin73,Baussch76} or
directly from the Fokker-Planck equation
\cite{Fogedby99a,Fogedby02a}, we have in the one dimensional case
analyzed the resulting coupled field equations minimizing the
action. The picture that emerges is that the transition
probabilities in the weak noise limit are associated with soliton
propagation and nucleation resulting from soliton collisions.

In this letter we apply a soliton approach in the canonical phase
space formulation to the noisy Ginzburg-Landau equation and
attempt to account for some of the numerical results of E et al.
We thus give analytical arguments for the propagation of noise
induced domain walls or solitons, the nucleation events associated
with domain wall creation and annihilation, and the associated
time dependent action. Details of our analysis will be given
elsewhere.

The noisy Ginzburg-Landau equation for a field $u(x,t)$ driven by
white noise has the form
\begin{eqnarray}
\frac{\partial u}{\partial t} = -\Gamma\frac{\delta F}{\delta u}
+\eta~,~~ \langle\eta(x,t)\eta(0,0)\rangle
=\Delta\delta(x)\delta(t), \label{gl}
\end{eqnarray}
with free energy
\begin{eqnarray}
F=\frac{1}{2}\int dx\left(\left(\frac{\partial u}{\partial
x}\right)^2+V(u)\right). \label{free}
\end{eqnarray}
In the switching problem considered by E et al. \cite{E02} $V(u)$
is given by the ``Mexican Hat''double well potential
\begin{eqnarray}
V(u) = k_0^2(1-u(x)^2)^2. \label{pot}
\end{eqnarray}
with strength parameter $k_0$. $\Gamma$ is a kinetic transport
coefficient setting the time scale.

The Ginzburg-Landau equation in its deterministic form has been
used both in the context of phase ordering kinetics \cite{Bray94}
and in its complex form in the study of pattern formation
\cite{Cross94}. In the noisy case for a finite system the equation
has been studied in \cite{Maier01}; see also an analysis of the
related $\phi^4$ theory in \cite{Habib00}. In the present problem
the noisy equation provides a generalization of the classical
Kramers problem \cite{Haenggi90} to spatially extended systems.

The equation admits a fluctuation-dissipation theorem yielding the
stationary distribution $P_{\text{stat}}\propto\exp[-2\Gamma
F/\Delta]$ . The equilibrium states follow from $\delta F/\delta
u=-d^2u/dx^2-2k_o^2u(1-u^2)=0$, giving the two degenerate uniform
ground states $u=\pm 1$ with $F=0$, as well as nonuniform domain
wall solutions
\begin{eqnarray}
u_{\text{dw}}(x)=\pm\tanh k_0(x-x_0), \label{dw}
\end{eqnarray}
centered at  $x_0$, connecting the two ground states. The
associated free energy is
\begin{eqnarray}
F_{\text{dw}}=4k_0/3. \label{freedomain}
\end{eqnarray}
Since the spectrum of $\delta^2F/\delta u^2=-d^2/dx^2 -
2k_0^2(1-3u^2)$ is positive for the ground states and has
zero-eigenvalue Goldstone modes (translation modes) for the domain
wall solutions, see e.g. \cite{Fogedby85}, we infer that the free
energy landscape possesses two global minima at $u=\pm 1$ and a
series of local metastable saddle points of free energy $4nk_0/3$,
corresponding to n connected domain walls at positions $x_i$,
$i=1,\cdots n$.

In order to address the issue of noise-driven transitions in the
Ginzburg-Landau equation (\ref{gl}) we apply the phase space
approach developed for the Kardar-Parisi-Zhang equation
\cite{Fogedby99a,Fogedby02a}. This method is based on a weak noise
WKB-like approximation,
$P(\{u(x)\},t)\propto\exp[-S(\{u(x)\},t)/\Delta]$, applied to the
Fokker-Planck equation $\Delta\partial P/\partial t = HP$ for the
transition probability $P(\{u(x)\},t)$, driven by the Hamiltonian
or Liouvillian $H(\{u(x)\},\{\partial/\partial u(x)\})$. To
leading order in $\Delta$ the action or weight function $S$ then
satifies the Hamilton-Jacobi equation $\partial S/\partial t +H=0$
with conjugate momentum or noise field $p(x)=\delta S/\delta
u(x)$. Moreover, the underlying principle of least action $\delta
S=0$ yields Hamiltonian equations of motion $\partial u/\partial
t=\delta H/\delta p(x)$ and $\partial p/\partial t=-\delta
H/\delta u(x)$. We note that the equations of motion are identical
to the saddle point equations in the Martin-Siggia-Rose functional
formulation \cite{Martin73,Baussch76}.

For the Ginzburg-Landau equation we then obtain explicitly the
coupled deterministic field equations
\begin{eqnarray}
&&\frac{\partial u}{\partial t} = \Gamma\frac{\partial^2
u}{\partial x^2}+ 2\Gamma k_0^2u(1-u^2) +p, \label{eq1}
\\
&&\frac{\partial p}{\partial t} = -\Gamma\frac{\partial^2 p}
{\partial x^2} -2\Gamma k_0^2p(1-3u^2), \label{eq2}
\end{eqnarray}
derived from the Hamiltonian
\begin{eqnarray}
H= \frac{1}{2}\int dx p\left(p+2\Gamma\frac{\partial^2u}{\partial
x^2}+4\Gamma k_0^2u(1-u^2)\right). \label{ham2}
\end{eqnarray}
In the weak noise limit the transition pathways are determined by
the orbits in the $(u,p)$ phase space from an initial
configuration $u_1(x)$ at time $t=0$ to a final configuration
$u_2(x)$ at time $t=T$. The conjugate field $p$ is a slaved
variable representing the noise driving the system. Finally, the
transition rate $P(u_1\rightarrow u_2,T)\propto\exp[-S/\Delta]$ is
determined by the action, $H=\int dx{\cal{H}}$,
\begin{eqnarray}
S(u_1\rightarrow
u_2,T)=\int_{u_1,0}^{u_2,T}dxdt\left[p\frac{\partial u}{\partial
t}-{\cal H}\right], \label{act1}
\end{eqnarray}
associated with the orbit.

To linear order, the effect of the noise field $p$ is to impart a
velocity to the static domain wall (\ref{dw}).  This is seen by
expanding $u$ and $p$ on the translation mode associated with the
static domain wall, $u_{\text{tm}}=-(1/m)du_{\text{dw}}/dx$, where
the mass $m$ is to be determined. Setting
$u=u_{\text{dw}}+u_0u_{\text{tm}}$ and $p=p_0u_{\text{tm}}$ and
using $(\delta^2F/\delta u^2)_{u_{\text{dw}}}u_{\text{tm}}=0$ we
obtain from (\ref{eq1}) and (\ref{eq2}) $du_0/dt=p_0$ and
$dp_0/dt=0$ with solutions $u_0=p_0t$ and $p_0=\text{const}$.
Consequently,
\begin{eqnarray}
u(x,t)\sim u_{\text{dw}}-\frac{p_0t}{m}\frac{u_{\text{dw}}}{dx}
\sim u_{\text{dw}}\left(x-\frac{p_0}{m}t\right),
\end{eqnarray}
describing a domain wall of mass $m$ propagating  with velocity
$v=p_0/m$. With the above normalization of the translation mode
the noise field $p_0$ is a momentum contributing to the total
momentum $\Pi=\int dx u\partial p/\partial x$ (generator of
translation)and thus canonically conjugate to the position $x$ of
the domain wall. From $\int dx\cosh^{-4}k_0x=4/3k_0$ we infer the
mass $m$. For the energy and action associated with the
propagation of a single domain wall in time $T$ we then have
\begin{eqnarray}
E_0=\frac{p_0^2}{2m},~~S_0=T\frac{p_0^2}{2m},~~ m=\frac{4k_0}{3}.
\end{eqnarray}
This analysis generalizes directly to a dilute gas of connected
non-overlapping domain walls. The noise $p$ then gives rise to
individual velocities imparted to each domain wall and the
energies, momenta, and actions simply add up.

In addition to the translation modes, time-dependent diffusive
modes are also excited, corresponding to small Gaussian
fluctuations about the local minima. In the neighborhood of $u=\pm
1$ it follows from (\ref{eq1}) and (\ref{eq2}) that $u$ and $p$ in
a plane wave decomposition, $\exp(ikx)$ evolve with
time-dependence $\exp(\pm\Gamma(k^2+4k_0^2)t)$. About the local
saddle points corresponding to domain wall propagation, the
extended modes are phase shifted corresponding to the trapping of
a localized deformation mode with time dependence $\exp(\pm\Gamma
3k_0^2t)$ and the translation mode discussed above, see e.g.
\cite{Fogedby85}.

The dynamical interpretation of the noise-induced switching in the
Ginzburg-Landau equation is now clear. The transition from e.g.
$u=+1$ to $u=-1$ starts by nucleating a small region of size $\sim
k_0^{-1}$ and then propagating a domain wall or walls, with
superposed linear modes subject to energy and momentum
conservation and topological constraints, until the whole system
is in the $u=-1$ state. The energy of the initial state is given
by $E=(1/2)\int dx p^2$, and the noise field thus has to be
assigned initially in order to reach the switched state $u=-1$ in
a prescribed time $T$. For topological reasons the domain walls
must nucleate and annihilate in pairs in general accompanied by
absorption or emission of linear modes. Since the linear modes
also carry positive action the dynamical modes with lowest action
correspond to nucleation or annihilation of domain wall pairs with
equal and opposite momenta, i.e., equal speeds.

In the case of periodic boundary conditions the momentum $\Pi=\int
dx u\partial p/\partial x$ of the initial and final states is
zero. The system is translational invariant and the formation and
annihilation of one or several domain wall pairs moving with the
same speed take place at equidistant positions along the axis. For
fixed boundary conditions the translational invariance is broken
and the momentum $\Pi$ is nonvanishing corresponding to nucleation
and annihilation of domain walls at the boundaries. This general
scenario of switching is completely consistent with the numerical
analysis in \cite{E02}.

Switching a system of size $L$ in time $T$ by means of a single
domain wall, corresponding to the pathway via the lowest local
saddle point of the free energy at $F_{\text{dw}}=m$, the
propagation velocity $v=p_0/m=L/T$ and we obtain the action
$S_1(T)=mL^2/2T$ and associated transition probability
$P\propto\exp(-mL^2/2\Delta T)$. In the thermodynamic limit
$L\rightarrow\infty$, $P\rightarrow 0$ as a result of the broken
symmetry in the double well potential. At long times the action
falls off as $1/T$. At intermediate times $t$ and positions $x$ we
have $P\propto\exp(-mx^2/2\Delta t)$ and we infer that the domain
walls in the stochastic interpretation perform a random walk with
mean square displacement $2\Delta t/m$, corresponding to diffusive
behavior.

\begin{figure}
\includegraphics[width=0.7\hsize]
{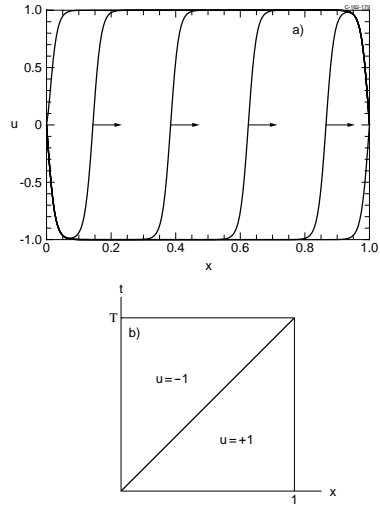} \caption{In a) the switching from $u=+1$ to $u=-1$ in
time $T$ is effectuated by means of a domain wall propagating with
velocity $v=1/T$. The domain wall is nucleated at $x=0$ and
annihilated at $x=1$. In b) the process is depicted in an $(x,t)$
plot.} \label{fig1}
\end{figure}

\begin{figure}
\includegraphics[width=0.7\hsize]
{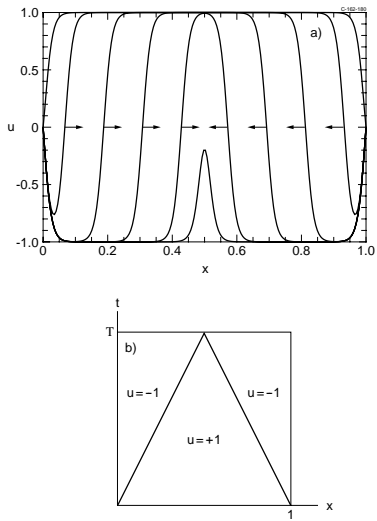} \caption { In a) the switching from $u=+1$ to $u=-1$ in
time $T$ takes place by means of two domain walls propagating in
opposite directions with velocity $v=1/2T$. The domain walls
nucleate at the boundaries and annihilate at the center. In b) the
switching process is depicted in an $(x,t)$ plot.} \label{fig2}
\end{figure}

\begin{figure}
\includegraphics[width=0.7\hsize]
{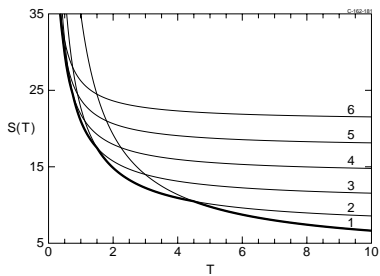} \caption{The action $S(T)$ given by (\ref{act2}) is
plotted as a function of $T$ for transition pathways involving up
to $n=6$ domain walls. The lowest action and thus the most
probable transition is associated with an increasing number of
domain walls at shorter times, indicated by the heavy limiting
curve. The curves correspond to choosing $S_{\text{nuc}}=5\Gamma
k_0$.} \label{fig3}
\end{figure}

In Fig.~\ref{fig1}a we show a domain wall nucleating at the left
boundary and propagating with constant velocity $v=1/T$ to the
right boundary, where it annihilates. We have used the same
parameter values as in \cite{E02}, i.e., $\delta = \Gamma = .03$,
$2\Gamma k_0^2= \delta^{-1}$, $T=7$, and a system size $L=1$.   In
Fig.~\ref{fig1}b we have plotted the trajectory of the domain wall
in space and time.
The switching can also take place by nucleating two domain walls
at the boundaries. These then move at half the velocity $v/2$ and
subsequently annihilate at the
center. This process corresponds to the pathway via the local
saddle point of the free energy at $F_{\text{dw}}=2m$, and the
action is given by $S_2(T)=2S_1(4T)$. Snapshots of this process are
shown in Fig.~\ref{fig2}a and the corresponding space-time plot
in Fig.~\ref{fig2}b.
Combining the contributions from nucleation and the subsequent
domain wall propagation, we can write a heuristic expression for
the total action:
\begin{eqnarray}
S_n(T)= nS_{\text{nuc}} + \frac{mL^2}{2nT}, \label{act2}
\end{eqnarray}
where $S_{\text{nuc}}$ is the action for nucleating a single
domain wall and $n$ is the number of walls. The action of
nucleation  is easily estimated from the Arrhenius factor
associated with the Kramers escape from  the ground state $u=+1$
to the saddle point in the free energy (\ref{free}), i.e.,
$P\propto \exp(-\Gamma F_{\text{dw}}/\Delta)$. We thus obtain a
nucleation action $S_{\text{nuc}}$ of order $\Gamma k_0$. A more
detailed argument, to be presented elsewhere, based on estimating
$S_{\text{nuc}}$ from the equations of motion (\ref{eq1}) and
(\ref{eq2}) and the action (\ref{act1}), yields for domain wall
nucleation and annihilation
\begin{eqnarray}
S_{\text{nuc}}\sim 6.5 \Gamma k_0. \label{snuc}
\end{eqnarray}
In Fig.~\ref{fig3} we have plotted $S$ versus $T$ for $n=1-6$
domain walls using the parameter values in \cite{E02}. Choosing
$S_{\text{nuc}}$ according to (\ref{snuc}) we find excellent
agreement with the numerical results.
As also discussed in \cite{E02} we note that the switching
scenario depends on $T$. At shorter switching times it becomes
more favorable to nucleate more domain walls. In the present
formulation this feature is associated with the finite nucleation
or annihilation action $S_{\text{nuc}}$. This is evidently a
finite size effect in the sense that the action at a fixed $T$
diverges in the thermodynamic limit $L\rightarrow\infty$,
corresponding to the broken symmetry.

In this letter we have presented a dynamical description and
analysis of a specific non equilibrium  transition in the noisy
Ginzburg-Landau equation based on a canonical phase space
formulation. We find good agreement both qualitatively and
quantitatively with the numerical finding of E. et al. \cite{E02}
based on an optimization of the Freidlin-Wentzel action. The
dynamical approach offers in the nonperturbative weak noise or low
temperature limit an alternative way of determining dynamical
pathways and the Arrhenius part of the associated transition
rates.
%

\end{document}